\documentclass[]{revtex4}
\usepackage{amsfonts,amssymb,amsbsy}
\usepackage{graphicx,amsfonts}
\usepackage{amsfonts,amssymb,amsbsy}
\usepackage{array}
\usepackage{amsmath}
\usepackage{amsfonts,amssymb,amsbsy}

\pdfoutput=1
\usepackage{graphicx,amsfonts}

\begin{document}

\title{The problems of $\eta'\rightarrow\pi^0\gamma\gamma$ decay and the New Physics
}

\author{Yaroslav Balytskyi}
\email{ybalytsk@uccs.edu}
\affiliation{Department of Physics and Energy Science, University of Colorado at Colorado Springs, Colorado Springs, Colorado 80933, USA
 }

\begin{abstract}

Rare decays of light mesons may be a discovery window for a new weakly coupled forces hidden at low energy QCD scale. BES-III Collaboration reported the observation of the rare decay $\eta'\rightarrow\pi^0\gamma\gamma$. The observed decay width  disagrees with the preliminary theoretical estimations. We show that this tension may be attributed to the New Physics, presumably Dark Photon. 

For the completeness, we consider the possible influence of the New Physics on a similar well-measured decay $\eta\rightarrow\pi^0\gamma\gamma$ and a recently measured one $\eta' \rightarrow \eta\gamma\gamma$ and show that the impact of the hypothetical Dark Photon may be also present in these decays also. 
 
\end{abstract}


\maketitle

\section{Introduction}
\label{intro}

Dark Photon is a hypothetical particle which should be a force carrier similar to photon in Electromagnetism and possibly connected to Dark Matter particles, and can be weakly coupled with the visible charged particles by a kinetic mixing with the usual photon, Ref.\cite{DarkPhotonCoupling}.
There is a number of anomalies which potentially could be caused by Dark Photon since it may be coupled to the usual photon. For instance, beryllium anomaly, Ref.\cite{Feng},  Dark Matter effects and astrophysics, Refs.\cite{AstroDM1, AstroDM2}, muon $(g-2)$, Refs.\cite{(g-2)1, (g-2)2, (g-2)3} and possibly "proton radius puzzle", Refs.\cite{ProtonRadius1, ProtonRadius2}.
Moreover, Dark Photon is a DM candidate itself, Ref.\cite{DM}. There is a Dark Photon  search carried out at JLab, Ref.\cite{JLab}, and at CERN, Ref.\cite{CERN}.

Nevertheless, in all these anomalies, searches and observations it is assumed that hypothetical Dark Photon has predominantly 
\textit{leptonic} coupling. On the contrary, in Ref.\cite{Tulin} was proposed a model of dark photon (or 
"$\mathcal{B}$ boson") which has both couplings to quarks and leptons and the coupling to quarks dominate over the coupling to leptons.

Moreover, as it was indicated in Ref.\cite{Tulin}, this area is not yet covered by the Beyond Standard Model searches. Consequently, rare decays $\eta'\rightarrow\pi^0\gamma\gamma$, $\eta'\rightarrow\eta\gamma\gamma$ and $\eta\rightarrow\pi^0\gamma\gamma$ could serve as a probe of such kind of Beyond Standard Model Physics. 

The proposed interaction Lagrangian has the form:

\begin{equation}
\mathcal{L}_{int}=(\frac{1}{3}g_B + \epsilon\cdot Q_q\cdot e)\cdot \bar{q}\gamma^{\mu}q\cdot\mathcal{B_{\mu}}- \epsilon\cdot e \cdot \bar{l}\gamma^{\mu} l \cdot \mathcal{B_{\mu}}
\end{equation}

 where $\epsilon$ - adjustable parameter. $\mathcal{B}$ boson mass was estimated in a range $140 \ MeV$ - $1 \ GeV$. It should have the same quantum numbers  as $\omega$ meson $I^G(J^{PC})=0^-(1^{--})$ to preserve the symmetries of low-energy QCD, Ref.\cite{Tulin}.

Dark Photon (or 
"$\mathcal{B}$ boson") could manifest itself as a resonance in rare decays of $\eta$, $\eta', \pi, \omega$ mesons, including $\eta^{\prime}\rightarrow\pi^0\gamma\gamma$ and similar $\eta\rightarrow\pi^0\gamma\gamma$, $\eta^{\prime}\rightarrow\eta\gamma\gamma$.

Since hypothetical Dark Photon may mix with the regular photon, it may also be coupled to all the three lightest vector mesons. However, the mixing with $\omega$ meson should be dominant. Photon is a linear combination of isoscalar and isovector and both $\omega$ meson and hypothetical Dark Photon are isoscalars, $\rho$ meson is isovector. $\phi(1020)$ is also isoscalar, however it gives a negligibly small contribution $\sim 1 \%$ to the width of these decays so the mixing of $\omega$ with Dark Photon should give the main contribution. Therefore, for purposes of our paper we neglect possible mixings of Dark Photon with vector mesons other than $\omega$.

Meanwhile, the branching ratio of $\eta'\rightarrow\pi^0\gamma\gamma$ decay  reported by BES-III collaboration  is, Ref.\cite{BES-III}:    

\begin{equation}
BR(\eta'\rightarrow\gamma\gamma\pi^0)_{Incl.}=(3.20 \pm 0.07 (stat) \pm 0.23(sys))\times 10^{-3}
\label{Branching}
\end{equation}

where the subscript \textit{"Incl."} means the branching fraction of the inclusive decay $\eta'\rightarrow\gamma\gamma\pi^0$.

From the theory side, the preliminary estimations  in the works Refs.\cite{Jora, Escribano, Ametller} were done using the combination of VMD (Vector Meson Dominance model), ChPT (Chiral Perturbation Theory) or $L\sigma M$ (Linear Sigma Model). 

\begin{figure}[h!]
  \includegraphics[width=15cm]{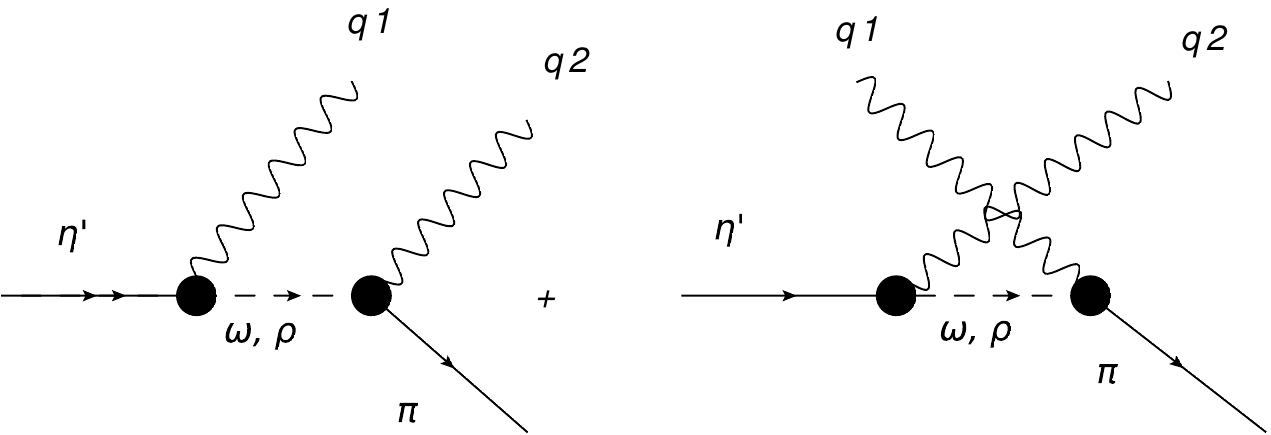}\\
  \caption{Leading order diagrams of $\eta'\rightarrow\pi^0\gamma\gamma$ decay}\label{Diagram}
\end{figure}

These estimations show that the decay is dominated by the intermediate vector mesons $\omega$ and $\rho$
subsequently decaying into $\pi^0\gamma$,   Fig.[\ref{Diagram}], and the decay width is estimated to be: $\Gamma_{\eta'\rightarrow\pi^0\gamma\gamma}=1.29 \  keV$, Ref.\cite{Escribano}, which is two times bigger the the observed result. 
Contributions both of the chiral loops and linear $\sigma$-terms are suppressed with respect to VMD on the level $\sim 10^{-3}$.

The VMD coupling constants are determined by $g$ (the vector-pseudoscalar-pseudoscalar coupling constant of
VMD), $\varphi_P$ ($\eta^{\prime}-\eta$ mixing angle) and $\varphi_V$ ($\omega-\phi$ mixing angle, which is zero if OZI rule is applied). All three quantities appear without any reference to the New Physics.

Nevertheless, the values of $g$, $\varphi_P$ and $\varphi_V$  extracted in different ways vary significantly, Ref.\cite{Phi1}, leading to a large scatter in the values of the coupling constants of vector mesons and, as a result, in the value of the theoretical branching ratio.

To reduce this uncertainty, in our calculation we used the values of the coupling constants of vector mesons extracted directly from the known decays. Extracted in such a way, their uncertainty is determined only by the uncertainties of the branching ratios of the particles, their widths and masses. \textit{However, as we will see, any choice  of coupling constants is unable to explain the discrepancy between the theoretical prediction and the experimental result.}

For the $L\sigma M$ contribution we use the value of the mixing angle extracted in our previous work, Ref.\cite{Mixing}. Note, in the $L\sigma M$ part the contribution of $a_0(980)$ was also taken into account with the amplitude taken from Ref.\cite{NewEscribano}.

We use a new parametrization of $\rho$ meson decay width on the energy, Ref.\cite{RhoWidthNew}, instead of the previously used, Ref.\cite{RhoWidth}, and take into account the OZI - suppressed $\phi (1020)$ - meson contribution. 

We find that the theoretical branching ratio of the decay lies in the range: 

\begin{equation}
BR^{VMD+L\sigma M} (\eta'\rightarrow\pi^0 \gamma \gamma)|_{Theoretical} = (7.88 \pm 1.26)\times 10^{-3}
\end{equation}

which is in a direct contradiction to the observed value, formula  (\ref{Branching}). 

There could be other possible contributions to this decay such as other possible intermediate states. However, their contribution should be small since they are heavy and thus are far from the boundaries of Dalitz plot. 

As we show, this discrepancy may be explained if we assume the existence of Dark Photon mixing with the ordinary $\omega$ meson without changing anything else in the model.

In addition, it's worth noting, recently an upper limit of the branching fraction of another decay $\eta'\rightarrow\eta\gamma\gamma$ was reported to be $1.33\times 10^{-4}$ at the 90$\%$ CL by BES-III Collaboration, Ref.\cite{BES-IIINEW}, which is again in a \textit{direct contradiction} with the theoretical prediction,  Ref.\cite{Escribano}. So, there are discrepancies in other similar rare decays which could be possibly attributed to Dark Photon effects.

Consequently, for the completeness of our analysis we consider a possible impact of $\mathcal{B}$ boson on the decay $\eta\rightarrow\pi^0\gamma\gamma$ in a similar manner with $\eta^{\prime}\rightarrow\pi^0\gamma\gamma$ since there is sufficient data available, Refs.\cite{etaUSUAL1, etaUSUAL2}, and show that its impact may be present in this decay also. However, if it's present, it should be smaller than for the case of $\eta^{\prime}\rightarrow\pi^0\gamma\gamma$.

For the case of $\eta^{\prime}\rightarrow\eta\gamma\gamma$ the fit is not possible yet since the invariant mass spectrum is not provided by BES-III, only the overall branching ratio and its upper limit are given. 

Nevertheless, we show that this tension can also be relaxed if we assume the existence of Dark Photon effectively changing the coupling constant of $\omega$ meson.

We postpone the joint fit of the parameters of $\mathcal{B}$ boson from these three decays simultaneously till more data on $\eta^{\prime}\rightarrow\eta\gamma\gamma$ becomes available, in particular, $\gamma\gamma$ spectrum.

\section{Theoretical predictions}\label{Prediction}
In the general case, the VMD amplitude of the decay corresponding to the diagram Fig.[\ref{Diagram}] is given by:
\begin{equation}\label{Amplitude}
A=(\frac{c_{\omega}}{D_{\omega}(t)}+\frac{c_{\rho}}{D_{\rho}(t)}+\frac{c_{\phi}}{D_{\phi}(t)})B(q_2)+
(\frac{c_{\omega}}{D_{\omega}(u)}+\frac{c_{\rho}}{D_{\rho}(u)}+\frac{c_{\phi}}{D_{\phi}(u)})B(q_1)
\end{equation}
 where $t=(P_{\eta'}-q_2)^2$, $u=(P_{\eta'}-q_1)^2$, $q_{1,2}$ - 4-momenta of outgoing photons, $P_{\eta'}$ - 4-momentum of $\eta'$ - meson, 
$D_{\omega, \rho}(t,u) = (t,u) - im_{\omega, \rho} \Gamma_{\omega, \rho} $ --  
propagator of vector meson (Breit - Wigner function). 
$B(q_{1,2})$ are kinematic coefficients representing the spin structure of the particles, Ref.\cite{Escribano}.

The VMD coupling constants are determined by 3 parameters: $g$, $\varphi_P$ and $\varphi_V$. In the limit of an exact OZI $\varphi_V = 0$, the coupling constants or $\omega$ and $\rho$ mesons are the same $c_{\omega}=c_{\rho}$ and determined by the pseudoscalar mixing angle $\varphi_P$, and for $\phi$ meson $c_{\phi}=0$.

\begin{math}
c^{OZI}_{\omega}=c^{OZI}_{\rho}=
(\frac{Ge}{\sqrt{2}g})^2\cdot \frac{1}{3}\cdot  Sin[{\varphi_P}], \ c_{\phi}^{OZI}=0 
\end{math}
, where $G=\frac{3 g^2}{4\pi^2 f_{\pi}}, g\approx 4.2$, $f_{\pi}$ - pion decay constant.

Nevertheless, the $\eta$-$\eta^{\prime}$ mixing angle is not uniquely defined.
We derived the mixing angle in our previous work, Ref.\cite{Mixing}, to be $\varphi_P=37.4^\circ \pm0.4 ^\circ$ from the analysis of charge
exchange reactions $\pi^{-}p$ and $K^{-}p$. 

In Ref.\cite{Phi1} the previous results on determination of the mixing angle data from different processes including strong
decays of tensor and higher-spin mesons, electromagnetic decays of vector and pseudoscalar mesons, $J/\psi$ decays into a vector and a pseudoscalar meson, and other transitions are summarized. They provide several values extracted in different ways: $\varphi_P=44.2^\circ\pm 1.4^\circ;
43.2^\circ\pm2.8 ^\circ; 40.7^\circ\pm3.7 ^\circ; 42.7^\circ\pm5.4^\circ; 41.0^\circ\pm3.5^\circ; 41.2^\circ\pm3.7^\circ; 50^\circ\pm26^\circ; 36.5^\circ\pm1.4^\circ; 
42.4^\circ\pm2.0^\circ; 40.2^\circ\pm2.8^\circ
$

Consequently, the coupling constants derived from the mixing angles can vary up to $\sim 30\%$ which can lead to a variation in the predicted decay width up to $\sim 50 \%$. 

Therefore, for the purposes of our paper, we derive the coupling constants directly from the known decays since the uncertainties in this case are related only to the masses of particles, their widths and branching ratios and are thus smaller. However, as we show further, \textit{for any choice of the coupling constants there is a discrepancy between the theoretical prediction and the experimental results}. 

In our approach, the constants of electromagnetic decays are
 $c_{\omega} = G_{\eta'\rightarrow \omega \gamma}\cdot G_{\omega\rightarrow \pi^{0} \gamma}$,
 $c_{\rho}= G_{\eta' \rightarrow \rho \gamma}\cdot G_{\rho \rightarrow \pi^{0} \gamma}$, $c_{\phi}=G_{\phi\rightarrow\pi^0\gamma}\cdot G_{\phi\rightarrow\eta'\gamma}$. They are determined from the known decay widths  $\eta' \rightarrow \omega \gamma$, $\eta' \rightarrow \rho \gamma$,
 $\omega \rightarrow \pi^{0} \gamma$, $\rho \rightarrow \pi^{0} \gamma$, $\phi\rightarrow\pi^0\gamma$ and $\phi\rightarrow\eta'\gamma$.

\begin{equation}
\Gamma(\omega\rightarrow\pi^0\gamma)=\frac{1}{3}\cdot G_{\omega\rightarrow\pi^0\gamma}^2 \cdot \frac{(m^2_{\omega}-m^2_{\pi^0})^3}{32\pi\cdot m^3_{\omega}}; \
\Gamma(\eta'\rightarrow\omega\gamma)=G_{\eta'\rightarrow\omega\gamma}^2 \cdot \frac{(m^2_{\eta'}-m^2_{\omega})^3}{32\pi\cdot m^3_{\eta'}}
\end{equation}

So $\omega$ coupling constant equals $ c_{\omega}^{\eta^{\prime}\rightarrow\pi^0\gamma\gamma}=G_{\omega\rightarrow\pi^0\gamma}\cdot G_{\eta'\rightarrow\omega\gamma}= ( 0.08872 \pm 0.00587 )  \ GeV^{-2}$ where the uncertainty is determined by the uncertainties of the masses of the particles, decay width and the branching ratio.

Analogously, for $c_\rho^{\eta^{\prime}\rightarrow\pi^0\gamma\gamma}$ coupling constant:

\begin{equation}
\Gamma(\rho\rightarrow\pi^0\gamma)=\frac{1}{3}\cdot G_{\rho\rightarrow\pi^0\gamma}^2 \cdot \frac{(m^2_{\rho}-m^2_{\pi^0})^3}{32\pi\cdot m^3_{\rho}}; \ 
\Gamma(\eta'\rightarrow\rho\gamma)= G_{\eta'\rightarrow\rho\gamma}^2 \cdot \frac{(m^2_{\eta'}-m^2_{\rho})^3}{32\pi\cdot m^3_{\eta'}}
\end{equation}

The corresponding decay constant is $c_{\rho}^{\eta^{\prime}\rightarrow\pi^0\gamma\gamma}=G_{\rho\rightarrow\pi^0\gamma} \cdot G_{\eta'\rightarrow\rho\gamma}= (
 0.08871 \pm 0.00892 )  \ GeV^{-2}$.

Finally, for $\phi$ meson: 

\begin{equation}
\Gamma(\phi\rightarrow\pi^0\gamma)= \frac{1}{3}\cdot G_{\phi\rightarrow\pi^0\gamma}^2 \cdot \frac{(m^2_{\phi}-m^2_{\pi^0})^3}{32\pi\cdot m^3_{\phi}}; \
\Gamma(\phi\rightarrow\eta'\gamma)= \frac{1}{3}\cdot G_{\phi\rightarrow\eta'\gamma}^2 \cdot \frac{(m^2_{\phi}-m^2_{\eta'})^3}{32\pi\cdot m^3_{\phi}}
\end{equation}

So, $c_{\phi}^{\eta^{\prime}\rightarrow\pi^0\gamma\gamma}=G_{\phi\rightarrow\pi^0\gamma} \cdot G_{\phi\rightarrow\eta'\gamma} = ( 0.00879 \pm  0.00036)\ GeV^{-2}$. Such a small value in comparison with $c_{\omega}^{\eta^{\prime}\rightarrow\pi^0\gamma\gamma}$ and  $c_{\rho}^{\eta^{\prime}\rightarrow\pi^0\gamma\gamma}$ is due to OZI. The coupling constants $c_{\omega}^{\eta^{\prime}\rightarrow\pi^0\gamma\gamma}$ and $c_{\rho}^{\eta^{\prime}\rightarrow\pi^0\gamma\gamma}$ in our approach are  approximately the same, and the difference between them is of order $\sim 10 \%$, $c_{\phi}$ is small in comparison with $c_{\rho}$ and $c_{\omega}$ but nonzero.

For similar decays $\eta^{\prime}\rightarrow\eta\gamma\gamma$, $\eta\rightarrow\pi^0\gamma\gamma$ the coupling constants extracted in a similar way from the known decays are:

\begin{equation}
\begin{cases}
c_{\omega}^{\eta\rightarrow\pi^0\gamma\gamma} = (0.09435 \pm 0.00641)\ GeV^{-2}\\
c_{\rho}^{\eta\rightarrow\pi^0\gamma\gamma} = (0.10718 \pm 0.01138)\ GeV^{-2} \\
c_{\phi}^{\eta\rightarrow\pi^0\gamma\gamma} = (0.00852 \pm 0.00027)\ GeV^{-2}
\end{cases};
\begin{cases}
c_{\omega}^{\eta^{\prime}\rightarrow\eta\gamma\gamma} = (0.01707 \pm 0.00168)\ GeV^{-2}\\
c_{\rho}^{\eta^{\prime}\rightarrow\eta\gamma\gamma} = (0.19116 \pm 0.01388)\ GeV^{-2} \\
c_{\phi}^{\eta^{\prime}\rightarrow\eta\gamma\gamma} = (-0.04657 \pm 0.00140)\ GeV^{-2}
\end{cases}
\label{Others}
\end{equation}

The $\omega$ - meson is quite narrow and its peaks are clearly seen on a Dalitz plot, but $\rho$ - meson is much wider, so we have to include the corrections due to the dependence of the width of $\rho$ - meson on energy which is dictated by the unitarity conditions,  Ref.\cite{PionLoop}.

We use a new parametrization of the $\rho$-meson width, Ref.\cite{RhoWidthNew}, instead of the one used previously, Ref.\cite{RhoWidth}, since, as it was shown in Ref.\cite{RhoWidthNew}, it gives equally good or better fits to the CMD2, SND, and KLOE collaborations data:

\begin{equation}
\Gamma_{\rho}(s)=\Gamma_{\rho}\cdot\frac{m_{\rho}}{\sqrt{s}}\cdot(\frac{s-4 \cdot m^2_{\pi^+}}{m^2_{\rho}-4 \cdot m^2_{\pi^+}})^{\frac{3}{2}}\cdot \theta(s-4\cdot m^2_{\pi^+})
\end{equation}

The finite-widths $\rho$ meson effects were also considered in Ref.\cite{4lepton}.

The VMD contribution is split in the following way: $\Gamma^{VMD}_{total}=\Gamma^{VMD}_{\omega}+\Gamma^{VMD}_{\rho}+\underbrace{\Gamma^{VMD}_{\omega-\rho}}_{Interference \ \omega - \rho}$. Their relative contributions are the following:
$\frac{\Gamma^{VMD}_{\omega}}{\Gamma^{VMD}_{total}}\approx 75 \%$, 
$\frac{\Gamma^{VMD}_{\rho}}{\Gamma^{VMD}_{total}}\approx 5 \%$,
$\frac{\Gamma^{VMD}_{\omega-\rho}}{\Gamma^{VMD}_{total}}\approx 20 \%$. Such a sharp difference between the contributions of $\rho$ and $\omega$ is due to the fact that $\Gamma_{\rho}\gg \Gamma_{\omega}$.
Nevertheless, the interference term is crucial in the area of a Dalitz plot outside the range of $\omega$  meson. 

\begin{figure}
\includegraphics[width=8cm]{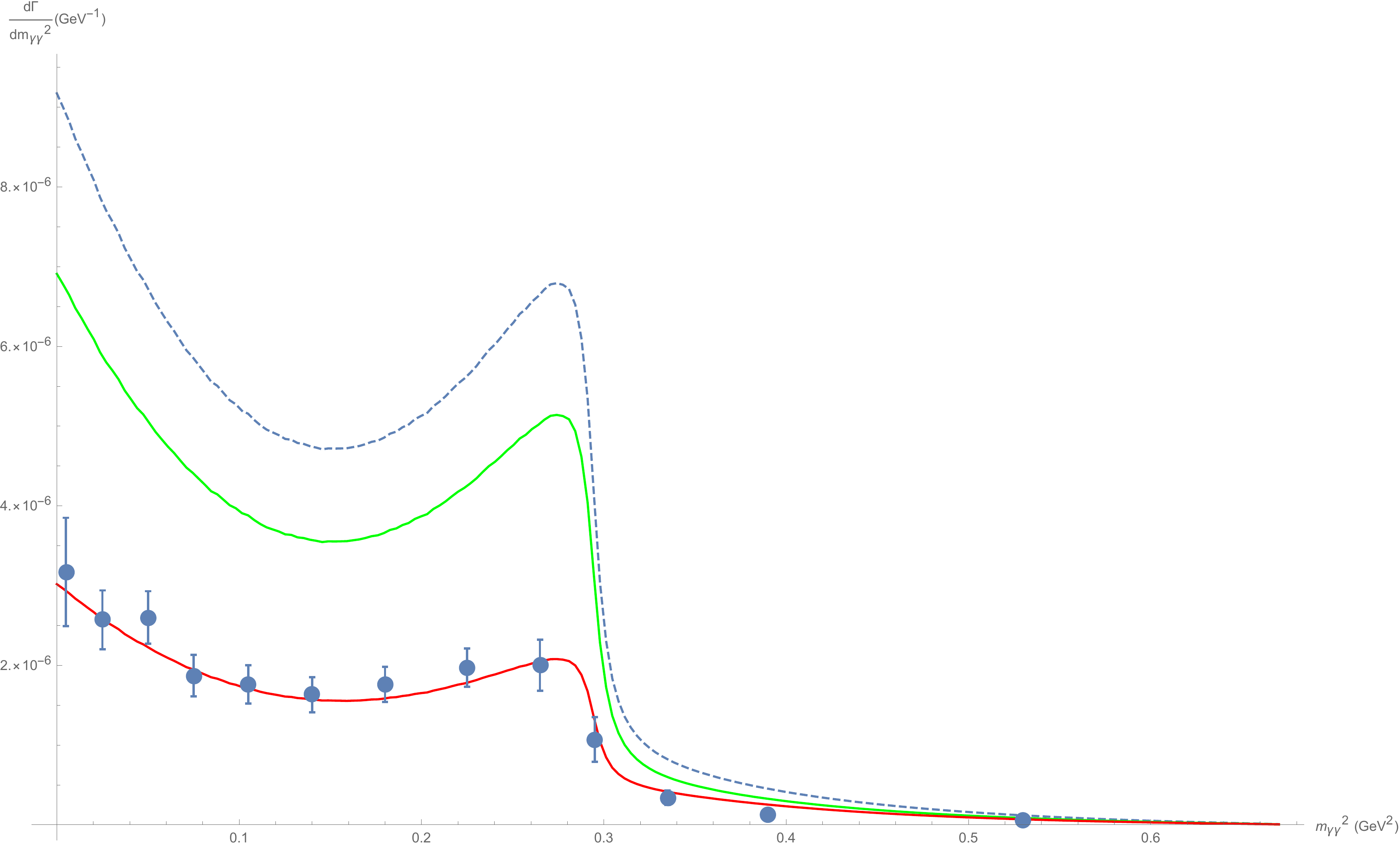}
\caption{$\frac{d\Gamma^{VMD+L\sigma M}_{\eta'\rightarrow\pi^0\gamma\gamma}}{dm^2_{\gamma\gamma}}$ compared to BES-III experimental points.  The upper blue dashed and the green lines correspond to the maximum and minimum values of the VMD coupling constants, the lower red line includes possible contribution of "$\mathcal{B}$ boson".}
\label{GammaGamma}
\end{figure}

We also include the contributions  of the kaon loops and $a_0(980)$ resonance in our calculation with the mixing angle determined in our previous work, Ref.\cite{Mixing}.  The corresponding spectrum $\frac{d\Gamma^{VMD+L\sigma M}_{\eta'\rightarrow\pi^0\gamma\gamma}}{dm^2_{\gamma\gamma}}$ is shown on Fig.[\ref{GammaGamma}].

The VMD contribution to the decay width is:

\begin{equation}
BR^{VMD} (\eta'\rightarrow\pi^0 \gamma \gamma)_{Theory} = (8.23 \pm 1.16) \times 10^{-3}
\end{equation}

The total decay width taking into account the coherent sum of VMD, kaon loops and $a_0(980)$ resonance: 

\begin{equation}
BR^{VMD+L\sigma M} (\eta'\rightarrow\pi^0 \gamma \gamma)_{Theory} = (7.88 \pm 1.26) \times 10^{-3}
\end{equation}

which is in a tension with the experimental result, formula (\ref{Branching}).

After the appearance of our manuscript, the new analysis of this decay, Ref.\cite{NewEscribano}, appeared. The authors of this work carried out a detailed and comprehensive fit of $\eta'\rightarrow\pi^0\gamma\gamma$, $\eta'\rightarrow\eta\gamma\gamma$ and $\eta\rightarrow\pi^0\gamma\gamma$ decays. In their approach, they simultaneously adjust three parameters: $\varphi_P$, $\varphi_V$ and $|g|$. In general, Ref.\cite{NewEscribano} confirms our conclusion about the discrepancy between the theoretical prediction and the experimental data and our predictions are close, the difference is due to different VMD coupling constants, mixing angles  and parametrizations of $\rho$ meson width used.

Nevertheless, in their approach they have \textit{three fitting parameters} and simultaneously fit \textit{three} coupling constants. 

In our approach, on the contrary, \textit{we don't have free parameters}.  To extract the values of the VMD coupling constants we used the known masses, decay widths and branching ratios of the particles. Moreover, in Ref.\cite{NewEscribano} the origin of the discrepancy was not determined. 

In the next section we discuss other possible contributions which could be the reason of these discrepancies.

In addition to the disrepancy in the decay we considered, $\eta^{\prime}\rightarrow\pi^0\gamma\gamma$, recently BES-III, Ref.\cite{BES-IIINEW}, provided the results for a similar decay $\eta'\rightarrow\gamma\gamma\eta$ which has a clear tension with theoretical prediction, Ref.\cite{Escribano}. 

Therefore, in addition to $\eta^{\prime}\rightarrow\pi^0\gamma\gamma$ it's worth considering the possible impact of hypothetical Dark Photon on similar decays $\eta\rightarrow\pi^0\gamma\gamma$ and $\eta^{\prime}\rightarrow\eta\gamma\gamma$.

\section{Other possible contributions}
As we seen, $\phi(1020)$ gives $\sim 1\%$ contribution to the overall decay width of $\eta^{\prime}\rightarrow\pi^0\gamma\gamma$. Also, as it was first shown in Ref.\cite{Escribano}, scalar meson effects of $a_0(980)$, $f_0(980)$ and $\sigma(600)$ which are included in $L\sigma M$ contribution give negligibly small contribution to all three decays.

Other possible intermediate states which could give the contribution to this decay can be found by constructing the invariant amplitudes. There are such options:
$
0^{-+}\rightarrow \gamma^{--}+1^{--} 
$, $
0^{-+}\rightarrow \gamma^{--}+1^{+-} 
$, $
0^{-+}\rightarrow \gamma^{--}+2^{+-} 
$\label{Options}.

Additional intermediate vector states could be: $\omega(1420)$, $\rho(1450)$, $\rho(1570)$, $\omega(1650)$, $\phi(1680)$, $\rho(1700)$, $\rho(1900)$, $\rho(2150)$, $\phi(2170)$. 
Possible axial intermediate states are: $h_1(1150)$, $b_1(1235)$, $h_1(1380)$. 

However, all these additional intermediate states are even heavier than $\phi(1020)$ and the aforementioned scalar mesons, thus they are even further from the boundaries of Dalitz plot. Therefore, neglecting them seems a safe assumption, and it's very unlikely that they could explain such a large discrepancy.

Another opportunity which could provide explanation of this discrepancy is Dark Photon (or "$\mathcal{B}$ boson"). On an experimental $\pi^0\gamma$ invariant mass spectrum the clear sharp peak of a new particle is not seen, Ref.\cite{BES-III}. 

Nevertheless, since "$\mathcal{B}$ boson" should have the same quantum numbers as $\omega$ meson, it can have mixing with $\omega$ meson and thus give a significant contribution to this decay.

If a possible contribution of "$\mathcal{B}$ boson" is taken into account, the formula (\ref{Amplitude}) should be modified in such a way: 

\begin{equation}
A \rightarrow (\frac{c_{\omega}}{D_{\omega}(t)}+\frac{c_{\rho}}{D_{\rho}(t)}+\frac{c_{\phi}}{D_{\phi}(t)}+\frac{c_{\mathcal{B}}}{D_{\mathcal{B}}(t)})B(q_2)+
(\frac{c_{\omega}}{D_{\omega}(u)}+\frac{c_{\rho}}{D_{\rho}(u)}+\frac{c_{\phi}}{D_{\phi}(u)}+\frac{c_{\mathcal{B}}}{D_{\mathcal{B}}(u)})B(q_1)
\end{equation}

For the purposes of this paper, we provide one set of parameters which can explain our discrepancy to show a viability  of such scenario. The exploring of the whole  parameter space of $\mathcal{B}$ boson we refer to our future work.

If we assume that the new $\mathcal B$ boson is hidden within the range of $\omega$ meson (so $m_{\mathcal{B}}=m_{\omega}$), then it's peak is not seen on the $\pi^0\gamma$ invariant mass spectrum. Nevertheless, it could give a significant contribution to the overall $\pi^0\gamma\gamma$ decay width. For instance, if the hypothetical $\mathcal{B}$ has a width such that $\Gamma_{\mathcal{B}}=\Gamma_{\omega}$ and the coupling constant $c_{\mathcal{B}}$ has an opposite sign than $c_{\omega}$, than the effective $\omega$ coupling constant would be lower than $c_{\omega}$: $c_{\omega}^{Effective}= c_{\omega} - |c_{\mathcal{B}}|<c_{\omega}$

As it was indicated in Ref.\cite{NewEscribano}, the BES-III result may be explained if we decrease the \textit{overall} normalization, so \textit{simultaneously} decrease $c_{\omega}$, $c_{\rho}$ and $c_{\phi}$ constants by decreasing $|g|$ and $Sin[\phi_P]$.

On the contrary, we introduce the effective $\omega$ meson coupling constant which may be caused by the possible mixing with "$\mathcal{B}$ boson"
\textit{without touching anything else} ($\rho$, $\phi$ mesons, kaon loops and $a_0(980)$).

Since this decay is dominated by the $\omega$ meson ($\sim 80\%$), it's decay widths is very sensitive to $\omega$ coupling constant. For our numerical estimations we take $c_{\omega}^{Effective} = 0.48\ GeV^{-2}$ and receive quite a similar result to Ref.\cite{NewEscribano} shown on Fig.[\ref{GammaGamma}] by a lower red line.

Consequently, assuming the BES-III result on $\eta^{\prime}\rightarrow\pi^0\gamma\gamma$ is correct, the scenario with the hypothetical Dark Photon (or $\mathcal{B}$ meson) is quite viable. 

Additionally, a recent measurement of $\eta^{\prime}\rightarrow\eta\gamma\gamma$ by BES-III Collaboration gives an upper limit on the branching ratio  $1.33\times 10^{-4}$ at the 90$\%$ CL , Ref.\cite{BES-IIINEW}, which is also in tension with the theoretical prediction,  Ref.\cite{Escribano}.

Consequently, the possible impact of hypothetical Dark Photon may be present in other similar rare decays $\eta^{\prime}\rightarrow\eta\gamma\gamma$ and $\eta\rightarrow\pi^0\gamma\gamma$ and we would like our model of Dark Photon to be flexible enough to explain the results of all three decays $\eta^{\prime}\rightarrow\pi^0\gamma\gamma$, $\eta^{\prime}\rightarrow\eta\gamma\gamma$ and $\eta\rightarrow\pi^0\gamma\gamma$ simultaneously.

Analogously to a regular $\omega$ meson, Dark Photon should have couplings to $\eta, \eta^{\prime}$ and $\pi^0$ and in each of those decays, like for a regular $\omega$ meson, the corresponding coupling constants should be determined as:
$c^{\eta^{\prime}\rightarrow\pi^0\gamma\gamma}_{\mathcal{B}} = G_{\mathcal{B}\eta^{\prime}\gamma}\cdot G_{\mathcal{B}\pi^0\gamma}$, $c^{\eta^{\prime}\rightarrow\eta\gamma\gamma}_{\mathcal{B}} = G_{\mathcal{B}\eta^{\prime}\gamma}\cdot G_{\mathcal{B}\eta\gamma}$, $c^{\eta\rightarrow\pi^0\gamma\gamma}_{\mathcal{B}} = G_{\mathcal{B}\eta\gamma}\cdot G_{\mathcal{B}\pi^0\gamma}$.

The measured branching ratio is less than theoretical prediction for $\eta^{\prime}\rightarrow\pi^0\gamma\gamma$ decay, bigger for $\eta\rightarrow\pi^0\gamma\gamma$ and smaller for
$\eta^{\prime}\rightarrow\eta\gamma\gamma$. This leads to the assumption that $G_{\mathcal{B}\eta^{\prime}\gamma}>0$, $G_{\mathcal{B}\eta\gamma}<0$, and $G_{\mathcal{B}\pi^0\gamma}<0$, therefore the "effective $\omega$ couplings" will be modified: 
\begin{equation}
\begin{cases} 
c^{\eta^{\prime}\rightarrow\pi^0\gamma\gamma}_{\mathcal{B}}<0;\Rightarrow c^{\eta^{\prime}\rightarrow\pi^0\gamma\gamma}_{\omega, Effective}< c^{\eta^{\prime}\rightarrow\pi^0\gamma\gamma}_{\omega}\\
c^{\eta\rightarrow\pi^0\gamma\gamma}_{\mathcal{B}}>0;\Rightarrow  c^{\eta\rightarrow\pi^0\gamma\gamma}_{\omega, Effective} > c^{\eta\rightarrow\pi^0\gamma\gamma}_{\omega}   \\
c^{\eta^{\prime}\rightarrow\eta\gamma\gamma}_{\mathcal{B}}<0; \Rightarrow c^{\eta^{\prime}\rightarrow\eta\gamma\gamma}_{\omega, Effective} < c^{\eta^{\prime}\rightarrow\eta\gamma\gamma}_{\omega}
\end{cases}
\label{Signs}
\end{equation}

Therefore, for $\eta'\rightarrow\pi^0\gamma\gamma$  and $\eta'\rightarrow\eta\gamma\gamma$ "effective $\omega$ coupling" would be reduced and for $\eta\rightarrow\pi^0\gamma\gamma$ it would be increased.

For the completeness, we consider a well-studied decay $\eta\rightarrow\pi^0\gamma\gamma$ in a similar manner to $\eta^{\prime}\rightarrow\pi^0\gamma\gamma$, and get the value which is \textit{smaller} than then the experimental result, Refs.\cite{etaUSUAL1, etaUSUAL2}:

\begin{equation}
BR^{VMD+L\sigma M} (\eta\rightarrow\pi^0 \gamma \gamma)|_{Theoretical} = (1.30 \pm 0.23)  \times 10^{-4} 
\end{equation}

\begin{equation}
BR^{VMD+L\sigma M} (\eta\rightarrow\pi^0 \gamma \gamma)|_{Experimental} = (2.56 \pm 0.22)  \times 10^{-4}
\end{equation}

Our result is \textit{smaller}
in comparison with theoretical prediction given in Ref.\cite{NewEscribano} since we used the coupling constants extracted directly from the known decays (and they are \textit{significantly smaller}), another parametrization for the $\rho$ meson width and value of the mixing angle. And, as in the case of $\eta'\rightarrow\pi^0\gamma\gamma$, we don't have free-fit parameters.

As it can be seen on Fig.[2] of Refs.\cite{NewEscribano}, the theoretical curve, although being on the edge of the error bars, lies noticeably lower than the majority of experimental points. Consequently, both our prediction and the prediction provided in Ref.\cite{NewEscribano} are \textit{smaller} than the experimental value, Refs.\cite{etaUSUAL1, etaUSUAL2}. 

\begin{figure}
\includegraphics[width=8cm]{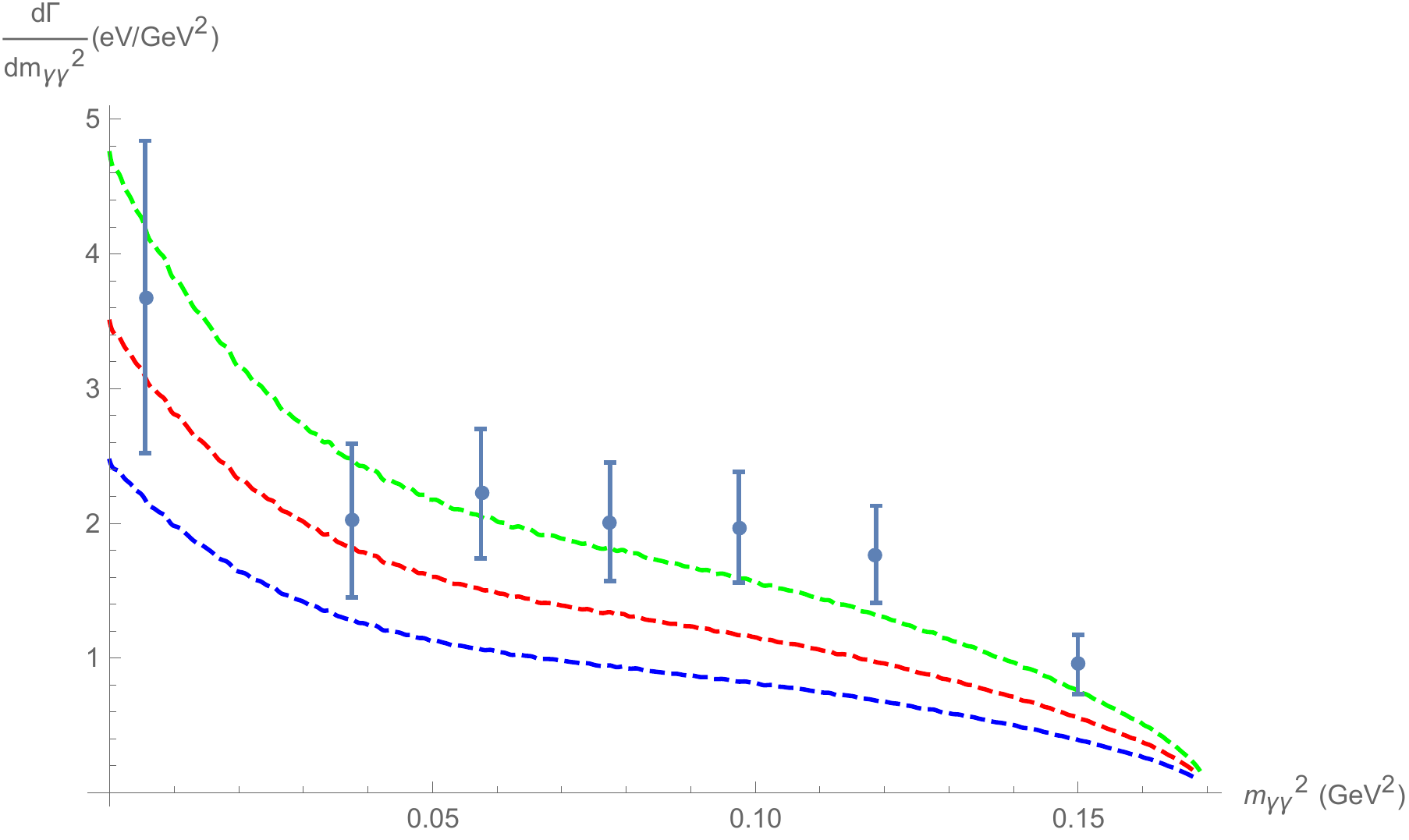}
\caption{$\gamma\gamma$ spectrum of $\eta\rightarrow\pi^0\gamma\gamma$. Two lower lines correspond to the largest and smallest values of the coupling constants. The upper green dashed line corresponds to a possible contribution of hypothetical $\mathcal{B}$ boson. The datapoints taken from Ref.\cite{etaUSUAL2}}
\label{etaREGULARtoPI0}
\end{figure}

Nevertheless, this discrepancy can be relaxed in a similar manner to $\eta'\rightarrow\pi^0\gamma\gamma$ if we assume the coupling constants of $\mathcal{B}$ to have the signs provided in formula (\ref{Signs}), so by increasing the effective $\omega$ coupling. For our numerical estimations, we take $c^{\eta\rightarrow\pi^0\gamma\gamma}_{\mathcal{B}} = 0.5\cdot c^{\eta\rightarrow\pi^0\gamma\gamma}_{\omega}$. For the decay $\eta\rightarrow\pi^0\gamma\gamma$ the effect of increasing the effective $\omega$ coupling on the $\gamma\gamma$ spectrum is shown on Fig.[\ref{etaREGULARtoPI0}]. 

The $\eta\rightarrow\pi^0\gamma\gamma$ decay is not dominated by $\omega$ meson like $\eta^{\prime}\rightarrow\pi^0\gamma\gamma$, so the effects of $\mathcal{B}$ meson distort the shape of $\gamma\gamma$ spectrum, not just shift it up. Nevertheless, as it can be seen on Fig.[\ref{etaREGULARtoPI0}],  the uncertainty bars are quite large, and the spectrum with the contribution of $\mathcal{B}$ boson approaches the allowed experimental values and the overall branching is increased approaching the experimental value.

Finally, for the case of $\eta^{\prime}\rightarrow\eta\gamma\gamma$ the direct comparison between theory and experiment is not possible yet since the $\gamma\gamma$ spectrum is not provided for this decay now. Nevertheless, it's clear that theoretical branching ratio of this decay may be decreased in a similar manner to $\eta'\rightarrow\pi^0\gamma\gamma$ if we assume the signs of the coupling constants of $\mathcal{B}$ boson to be given by formula (\ref{Signs}), since in the case $\eta'\rightarrow\eta\gamma\gamma$ the effective $\omega$ coupling constant has to be reduced. 

For example, taking into account that  $c^{\eta\rightarrow\pi^0\gamma\gamma}_{\omega}\approx 
c^{\eta^{\prime}\rightarrow\pi^0\gamma\gamma}_{\omega}$:

\begin{equation}
\begin{cases}
c^{\eta^{\prime}\rightarrow\pi^0\gamma\gamma}_{\mathcal{B}} = G_{\mathcal{B}\eta^{\prime}\gamma}\cdot G_{\mathcal{B}\pi^0\gamma}\approx -0.5\cdot c^{\eta^{\prime}\rightarrow\pi^0\gamma\gamma}_{\omega}\\
  c^{\eta\rightarrow\pi^0\gamma\gamma}_{\mathcal{B}} = G_{\mathcal{B}\eta\gamma}\cdot G_{\mathcal{B}\pi^0\gamma} \approx 0.5 \cdot c^{\eta\rightarrow\pi^0\gamma\gamma}_{\omega} \\
  c^{\eta^{\prime}\rightarrow\eta\gamma\gamma}_{\mathcal{B}} = G_{\mathcal{B}\eta^{\prime}\gamma}\cdot G_{\mathcal{B}\eta\gamma} ,
\end{cases}\Rightarrow
\frac{G_{\mathcal{B}\eta^{\prime}\gamma}}{G_{\mathcal{B}\eta\gamma}}\approx -1,   \ c^{\eta^{\prime}\rightarrow\eta\gamma\gamma}_{\mathcal{B}} \approx -(G_{\mathcal{B}\eta^{\prime}\gamma})^2 
\label{EtaPrimeEta}
\end{equation}

Taking $(G_{\mathcal{B}\eta\gamma})^2 \approx  c^{\eta^{\prime}\rightarrow\eta\gamma\gamma}_{\omega}$ and knowing the relative contributions of $\omega$, $\rho$ and $\phi$ we can make $c^{\eta^{\prime}\rightarrow\eta\gamma\gamma, \ Effective}_{\omega}\approx 0$.
Consequently for the case of the $\eta^{\prime} \rightarrow\eta\gamma\gamma$ decay, the theoretical decay width may be reduced by $\sim 40\%$   in a similar manner to other two aforementioned decays thus reducing the tension with the experimental result.

We postpone the detailed analysis of $\eta^{\prime}\rightarrow\eta\gamma\gamma$ decay and the simultaneous fit of parameters of hypothetical $\mathcal{B}$ boson, $\{m_{\mathcal{B}}, \Gamma_{\mathcal{B}}, G_{\mathcal{B}\eta^{\prime}\gamma}, G_{\mathcal{B}\pi^0\gamma}, G_{\mathcal{B}\eta\gamma}\},$
till more experimental data becomes available, in particular, $\gamma\gamma$ spectrum of $\eta^{\prime}\rightarrow\eta\gamma\gamma$ decay.

\section{Conclusions}
For any choice of the coupling constants there is a clear discrepancy between $\Gamma(\eta^{\prime}\rightarrow\pi^0\gamma\gamma)^{VMD+L\sigma M}_{Theory}$ and the observed result by BES-III which can be attributed to the New Physics, presumably Dark Photon (or "$\mathcal{B}$ boson"). 

As may be seen on Fig.[\ref{GammaGamma}], the scenario with the $\mathcal{B}$ boson giving a contribution to $\eta^{\prime}\rightarrow\pi^0\gamma\gamma$ is quite viable.

Considering in a similar manner the decay $\eta\rightarrow\pi^0\gamma\gamma$ we see that both our approach, and the approach provided in Ref.\cite{NewEscribano} give the values which are  below the experimental result, Refs.\cite{etaUSUAL1, etaUSUAL2}. 

Unlike $\eta^{\prime}\rightarrow\pi^0\gamma\gamma$, $\eta\rightarrow\pi^0\gamma\gamma$ is not dominated by $\omega$ boson, so the shape of $\gamma\gamma$ spectrum is distorted. Nevertheless, taking into account that the error bars are large, there is more space for the change of spectrum shape by taking into account $\mathcal{B}$ boson contribution, Fig.[\ref{etaREGULARtoPI0}].

Additionally, a recent measurement of $\eta^{\prime}\rightarrow\eta\gamma\gamma$, Ref.\cite{BES-IIINEW} gives the branching ratio which is also significantly smaller than the theoretical prediction,  Ref.\cite{Escribano}. Clearly, by reducing the "effective $\omega$ coupling" we can reduce the tension between theory and experiment in a similar manner to $\eta^{\prime}\rightarrow\pi^0\gamma\gamma$ in this case also. However, we postpone this analysis till more data on this decay becomes available.

\section{Acknowledgements}
Y.B. appreciates the help of A. Lykhodid and useful discussions with V. Samoylenko, V.Kiselev, A. Pinchuk, A. Zaitsev and S. Sadovsky.

\end{document}